\documentstyle[twocolumn,sproc]{article}
\newcommand{\Od}{{\cal O}}
\newcommand{\tr}{\mbox{tr}}

\newcommand{\NP}[1]{{\em Nucl.\ Phys.\ }{\bf #1}}
\newcommand{\ZP}[1]{{\em Z.\ Phys.\ }{\bf #1}}

\newcommand{\PL}[1]{{\em Phys.\ Lett.\ }{\bf #1}}

\begin{document}

%
%

\input epsf


{\onecolumn 

\hfill LBNL-39251\\

\vspace{3cm}
\begin{center}
{\LARGE \bf Strong WW Scattering\\
Chiral Lagrangians, Unitarity and Resonances}
\footnotemark[1]\\
\vspace{1.5cm}
{\Large
J. R. Pel\'aez\footnotemark[2]\\ {\it Theoretical Physics Group}\\
{\it Ernest Orlando Lawrence Berkeley National Laboratory}\\
{\it University of California, Berkeley, California  94720}}
\end{center}

\footnotetext[1]
{This work was supported by the Director, Office of Energy
Research, Office of High Energy and Nuclear Physics, Division of High
Energy Physics of the U.S. Department of Energy under Contract
DE-AC03-76SF00098.}

\footnotetext[2]{Complutense del Amo fellow. 
On leave of absence from
Departamento de F\'{\i}sica Te\'orica. Univ. Complutense.
28040 Madrid, Spain.}

\vskip 1.cm

\leftskip 1cm
\rightskip 1cm
\begin{abstract}
{\large 
Chiral lagrangians provide a model independent description of the
strongly interacting symmetry breaking sector. 
In this work it is first reviewed 
the LHC sensitivity to the chiral parameters
(in the hardest case of non-resonant low-energy WW scattering).
Later it is shown 
how to reproduce or predict the resonance spectrum by means 
of dispersion theory and the inverse amplitude method.
We present a parameter space scan that covers many different
strong WW scattering scenarios.}
\end{abstract}

\vskip 1.5cm

\begin{center}
{\large
 Talk presented at the: \\ 1996 DPF/DPB
Summer Study on New Directions for High-Energy Physics.\\
(Snowmass 96). }
\end{center}

\leftskip 0cm
\rightskip 0cm

\newpage

\vspace*{3.in}

\begin{center}
{\bf Disclaimer}
\end{center}

\vskip .2in

\begin{small}
\begin{quotation}
This document was prepared as an account of work sponsored by the United
States Government. While this document is believed to contain correct
 information, neither the United States Government nor any agency
thereof, nor The Regents of the University of California, nor any of their
employees, makes any warranty, express or implied, or assumes any legal
liability or responsibility for the accuracy, completeness, or usefulness
of any information, apparatus, product, or process disclosed, or represents
that its use would not infringe privately owned rights.  Reference herein
to any specific commercial products process, or service by its trade name,
trademark, manufacturer, or otherwise, does not necessarily constitute or
imply its endorsement, recommendation, or favoring by the United States
Government or any agency thereof, or The Regents of the University of
California.  The views and opinions of authors expressed herein do not
necessarily state or reflect those of the United States Government or any
agency thereof, or The Regents of the University of California.
\end{quotation}
\end{small}

\vskip 2in

\newpage

\title{Strong WW Scattering\\
Chiral Lagrangians, Unitarity and Resonances
\thanks{This work was supported by the Director, Office of Energy
Research, Office of High Energy and Nuclear Physics, Division of High
Energy Physics of the U.S. Department of Energy under Contract
DE-AC03-76SF00098.}}

\author{J. R. Pel\'aez\footnotemark[1]\\ {\it Theoretical Physics Group}\\
{\it Ernest Orlando Lawrence Berkeley National Laboratory}\\
{\it University of California, Berkeley, California  94720}}

\maketitle

\footnotetext[1]{Complutense del Amo fellow. 
On leave of absence from
Departamento de F\'{\i}sica Te\'orica. Univ. Complutense.
28040 Madrid, Spain.}

\thispagestyle{empty}\pagestyle{empty}

\begin{abstract} 
Chiral lagrangians provide a model independent description of the
strongly interacting symmetry breaking sector. 
In this work it is first reviewed 
the LHC sensitivity to the chiral parameters
(in the hardest case of non-resonant low-energy WW scattering).
Later it is shown 
how to reproduce or predict the resonance spectrum by means 
of dispersion theory and the inverse amplitude method.
We present a parameter space scan that covers many different
strong WW scattering scenarios.
\end{abstract}

\section{Chiral Lagrangians}
\subsection{Introduction}
In the Standard Model (SM)
 there is an spontaneous symmetry breaking of the
gauge $SU(2)_L\times U(1)_Y$ group down to $U(1)_{EM}$.
The underlying theory that produces this mechanism is
unknown to a large extent. 
Basically, what we know is the following:

\begin{Itemize}
\item There is a system with a global symmetry breaking
from a group $G$ down to another one $H$ producing 
three Goldstone bosons (GB). 
\item The scale 
of this new interactions is $v\simeq250\mbox{GeV}$. 
\item The electroweak $\rho$ parameter is very close to one.
\end{Itemize}
\noindent
This last requirement is most naturally satisfied if the 
electroweak Symmetry Breaking Sector (EWSBS) 
respects the so called 
custodial symmetry
$SU(2)_{L+R}$ \cite{Sikivie}. 
Demanding just three GB, we are 
lead to $G=SU(2)_L\times SU(2)_R$ and $H=SU(2)_{L+R}$
\cite{LET,NPB}.

That is the very same breaking pattern of chiral symmetry in 
QCD with two massless quarks. 
It is well known that a rescaled version
of QCD is not valid as an EWSBS.
However, we still can borrow 
the formalism of chiral lagrangians
\cite{DoHe}, 
 known as Chiral Perturbation Theory (ChPT),
which
works remarkably well for pion physics
\cite{GaLe}.

Our case is different to QCD since, 
among other things, the GB disappear in the Higgs
mechanism. They become the longitudinal components of
the gauge bosons. Hence, if we want to probe an
strong EWSBS, we actually have to
look at interactions of longitudinal gauge bosons.
(We will denote both $W$ and $Z$ by $V$).
Indeed if the EWSBS is
strongly interacting, we expect an enhancement in
$V_L$ production.
That is why we are interested in $V_LV_L$ scattering.

\subsection{The Low energy Theorems}
The chiral lagrangian is built as a (covariant) derivative 
expansion out of GB fields. Only those operators respecting
the above symmetry pattern and Lorentz invariance are allowed
(we are also neglecting CP violation).
Thus, there is only one possible term with two derivatives:
\begin{equation}
{\cal L}^{(2)}=\frac{v^2}{4}\tr D_\mu UD^\mu U^\dagger
\label{NLSM}
\end{equation}
where the GB fields $\pi^i$ are
collected in the $SU(2)$ matrix
$U=\exp(i\pi^i \sigma^i/v)$ and $D_\mu$ is 
the usual covariant derivative. 

The above lagrangian is able to describe the very low energy 
behavior of the EWSBS. However it will be useful when 
{\em only} the GB and the gauge fields are relevant 
at low energies. That is the case of the strong EWSBS 
since the other particles affecting $VV$ scattering 
(like resonances) are expected at the TeV scale.

It is important to remark that the lagrangian in Eq.1
only depends on the symmetry 
structure and the scale. Its predictions for $V_LV_L$ scattering
are therefore universal. The two derivatives become
external momenta and thus
this term yields $\Od(p^2)$ contributions, which are called
the Low Energy Theorems (LET) \cite{LET}.
\subsection{The $\Od(p^4)$ lagrangian.}
The lagrangian in Eq.\ref{NLSM}
is that of a non-linear $\sigma$ model. Thus, in a strict sense it is 
non-renormalizable. However, all the divergencies appearing at 
one loop are $\Od(p^4)$ and can be absorbed in the parameters 
of the ${\cal L}^{(4)}$ lagrangian. If we were to consider two
loops with ${\cal L}^{(2)}$ we would need the ${\cal L}^{(6)}$
lagrangian and so on. The relevant point is that up to a given 
order in the external momenta the calculations can be renormalized 
and are finite.

There are many terms in the ${\cal L}^{(4)}$ lagrangian \cite{Appel}, 
although for $VV$ scattering at $\Od(p^4)$
it is enough to consider:
\begin{eqnarray}
{\cal L}^{(4)}&=&
L_1 \left( \tr D_\mu UD^\mu U^\dagger \right)^2
+ L_2 \left( \tr D_\mu UD^\nu U^\dagger \right)^2\nonumber \\
&+&i \tr \left[
(L_{9L} W^{\mu\nu}+L_{9R} B^{\mu\nu}) 
D_\mu UD_\nu U^\dagger
\right]
\nonumber \\
&+& L_{10}\tr U^\dagger B^{\mu\nu} U W_{\mu\nu}
\label{L4}
\end{eqnarray}
where $W^{\mu\nu}$ and $B^{\mu\nu}$ are the strength tensors 
of the gauge fields.
Only the values of the $L_i$ parameters depend
on the underlying theory.

For our purposes, we are only interested in $L_2$ and
$L_1$, which are the ones that enter the $VV$ fusion calculations.
The others are related to anomalous couplings.
 Their values can be estimated for the
minimal SM (MSM) with a heavy Higgs \cite{MJEsther}
as well as for QCD-like models (using the
ChPT parameters \cite{IAM}).
In Table I we give some reference values. 
\begin{table}[h]
\begin{center}
\caption{Chiral Parameters for different reference models.}
\vskip .3cm
\begin{tabular}{|l|cc|}\hline
& $L_1$ & $L_2$\\ \hline
MSM  ($M_H\sim1$ TeV)& 0.007 & -0.002 \\
QCD-like & -0.001 & 0.001 \\ \hline
\end{tabular}
\end{center}
\end{table}
Notice that in the literature it is also 
frequent to extract a $16\pi^2$ factor
so that the $L_i$ are of order 
unity.

Using the lagrangians in Eqs.\ref{NLSM} and \ref{L4}
 we can calculate the $VV$ elastic scattering amplitudes.
Indeed they are obtained as a truncated series in $p/4\pi v$,
as follows:
\begin{equation}
t(s)\simeq t^{(0)}(s) + t^{(1)}(s) + \Od(p^6)
\label{trunc}
\end{equation}
Where $t^{(0)}(s)$ is $\Od(p^2)$ and reproduces the LET.
It is obtained from ${\cal L}^{(2)}$ at tree level.
The $t^{(1)}(s)$ contribution is $\Od(p^4)$ and
comes from the ${\cal L}^{(4)}$ at tree level and
${\cal L}^{(2)}$ at one loop. If we made one more loop
we would get $\Od(p^6)$ contributions, 
and we would need the ${\cal L}^{(6)}$ lagrangian, etc...

Note that a naive estimate of the applicability range
is $4\pi v\lappeq 3\mbox{TeV}$. However, the existence
of resonances will limit the effectiveness of the approach 
up to $\lappeq 1.5\mbox{TeV}$.

\subsection{Chiral parameters at LHC}

The goal of future accelerators is to determine the nature
of the EWSBS. As we have seen, chiral lagrangians provide a 
model independent formalism. We always deal with the same set of
operators and only the actual values of the parameters
depend on the fundamental theory. 

As we have already stressed the most natural channel
to look for strong EWSBS interactions is $V_LV_L$ scattering.
The most striking experimental feature
 would be the appearance of resonant states. 
However, it is not assured that they could be directly
seen in the next generation of colliders. Even though they
are expected at the TeV scale, they can be higher
that the planned energy reach. In that case one is left with
a non-resonant behavior, where different models
will be hard to distinguish. Then the
effective lagrangians become a natural and systematic
tool to parametrize and maybe disentangle
the experimental results.

Indeed there are already some studies of the capability
of LHC to measure the chiral parameters \cite{CMS}. In Table II
are listed the number of events 
produced with various non vanishing values of 
$L_2$ or $L_1$. Following reference \cite{CMS}
we have recalculated the results for 100fb$^{-1}$ of
integrated luminosity at $\sqrt{14} \mbox{TeV}$. 
That corresponds to one experiment
collecting data at full design luminosity during one year.

The numbers in Table II are those of the cleanest
leptonic decays of subprocesses whose final state 
is either $W^\pm Z$ or $ZZ$:
\begin{eqnarray*}
q\bar q'\rightarrow W^\pm Z &\quad&q\bar q\rightarrow Z Z  
\quad\quad\quad gg\rightarrow ZZ \\
W^\pm Z\rightarrow W^\pm Z&\quad& ZZ\rightarrow ZZ \\
W^\pm \gamma\rightarrow W^\pm Z &\quad& W^+W^-\rightarrow ZZ  \\
\end{eqnarray*}
They have been calculated from the lagrangian in Eqs.\ref{NLSM}
and \ref{L4} at tree level (except gluon fusion, that only 
occurs at one loop). All possible initial and final
helicity combinations have been considered. We use the effective W
approximation, but {\em not} the Equivalence Theorem.
\begin{table}[h]
\begin{center}
\caption{Number of events and statistical significances
for different values of $L_2$ and $L_1$ at LHC.}
\begin{tabular}{|l|c|c|c|c|}   \hline
 & $10^{-2}$ & -$10^{-2}$ &  $5\times10^{-3}$ & 
-$5\times10^{-3}$ \\  \cline{2-5}
 &  \multicolumn{4}{c|}{$L_1$} \\  \hline 
$W^{\pm}Z^0
\rightarrow W^{\pm}Z^0$  & 22 & 58 & 23 & 41 \\ 
total $W^{\pm}Z^0$ &  104 &  139 & 105 & 122 \\ \hline
${r_5 |}_{W^{\pm}Z^0}$ & 0.7 & 2.6 & 0.6 & 1.0 \\  
${r_5 |}_{W^{\pm}Z^0\; tagging}$ & 1.0 & 4.2 & 0.9 & 1.7 \\  \hline
$W^+W^-
\rightarrow  Z^0Z^0$ &  21  & 7 &  13 &  6 \\ 
$Z^0Z^0  \rightarrow  Z^0Z^0$ & 6 & 6 & 1 & 1 \\ 
total $Z^0Z^0$ & 46 & 32 & 33 & 26 \\ \hline
${r_5|}_{Z^0Z^0}$ & 3.8 & 0.9 & 1.2 & 0.1 \\ 
${r_5|}_{Z^0Z^0\; tagging}$ & 6.6 & 1.8 & 2.3 & 0.2 \\ \hline
\hline 
 & \multicolumn{4}{c|}{$L_2$} \\ \hline
$W^{\pm}Z^0   \rightarrow W^{\pm}Z^0$ & 36 & 80 & 27 & 47 \\ 
total $W^{\pm}Z^0$ & 118 & 162 & 109 & 129 \\ \hline
${r_4 |}_{W^{\pm}Z^0}$ & 0.7 & 4.8 & 0.2 & 1.7 \\ 
${r_4 |}_{W^{\pm}Z^0\; tagging}$ & 1.0 & 7.5 & 0.3 & 2.7 \\ \hline
$W^+W^-\rightarrow  Z^0Z^0$ &  12  & 7 &  9 &  7 \\ 
$Z^0Z^0  \rightarrow  Z^0Z^0$ &  6 & 6 & 1 & 1 \\ 
total $Z^0Z^0$ & 37  & 32 & 30 & 27 \\ \hline
${r_4|}_{Z^0Z^0}$ & 1.9 & 0.9 & 0.5 & $\simeq$0 \\ 
${r_4|}_{Z^0Z^0\; tagging}$ & 3.5 & 1.8 & 0.9 & 0.1 \\ \hline
\end{tabular}
\end{center}
\end{table}
By the cleanest leptonic modes we mean the $W$'s and the $Z$'s
decaying to $\nu_ee,\nu_\mu\mu$ and $e^-e^+,\mu^-\mu^+$, 
respectively. The corresponding branching ratios are BR(WZ)=0.013
and BR(ZZ)=0.0044. We have also imposed a set of minimal cuts:
$M^{\max}_{VV}=1.5\mbox{TeV}, P^Z_{T\; min}=300 \mbox{GeV}, y^V_{max}=2$.
Further details of the calculation can be found in \cite{CMS}.

The statistical significances are defined with respect to the
"zero" model (when all the $L_i$ are set to zero).
In \cite{CMS} they are also given 
with respect to the SM with $M_H\simeq 1\mbox{TeV}$.
Note that the zero model
 is nothing but the LET predictions or the 
$M_H\rightarrow \infty$ limit of the MSM. The statistical significances
are defined as:
\begin{equation}
r_i=\frac{\vert N(L_i)-N(0)\vert}{\sqrt{N(0)}}
\end{equation}

In Table II we have listed two sensitivities for each process
depending on whether there is forward jet tagging available or not.
This detector feature is very important to separate those events coming
from $VV$ fusion from those coming from quarks.
We have given numbers for no jet tagging
at all and 100\% efficiency tagging, so that the real number will 
lie somewhere in between.

The analysis is simplified in the sense that only one $L_i$
is different from zero at a time. However,
there are issues that could improve the sensitivity that we 
have not addressed. We have only restricted ourselves to
leptonic modes, and we have not studied the $W^+W^-$
or the $W^\pm W^\pm$ final states. The sensitivities only refer
to separate channels and a simultaneous fit to all them
would be a considerable improve. There is still open the
possibility of final state polarization analysis that would enhance the
longitudinal modes. Finally we are also confident that more elaborated
cuts will also enhance the signal. Therefore, we think that
the numbers in Table II can be considered
 as a conservative estimate of the LHC
capabilities.

From Table II we can thus see that the $10^{-2}$ values are
at hand at the 3$\sigma$ level, both for $L_2$ and
$L_1$. Combining the two experiments and one or two
years of running even the 5$\sigma$ level seems attainable.

It is convenient at this point
to look back at Table I and notice that the
expected values lie on the range $10^{-2}$ to $10^{-3}$.
Therefore, we can easily reach the beginning of the interesting region.
Notice also that the two reference models have different signs
in their parameters.
Fortunately the experimental signature
is radically different when changing the sign of the parameters.
It seems feasible to differentiate positive from negative signs.

To go down to the level of $L_i=5\times10^{-3}$
its harder, but not impossible. The 3$\sigma$ level seems reachable
in three or four years in many channels, by combining 
the two experiments. We have not listed  the results for $10^{-3}$
since that level of precision seems extremely hard to access
\cite{CMS}.

It is important to remark again that this is a preliminary
and conservative result. We can conclude that even in the non-resonant
scenario, LHC will be able to test at least {\em  part} of the chiral
parameter space in the interesting region.
It is also clear that the study of this kind of physics will require
the ultimate machine performance.

As we will see in the next section the determination of 
$L_1$ and $L_2$ will be very helpful to
disentangle the nature of an strong EWSBS. Even if the
LHC energy reach is not enough to observe resonances directly,
their existence can be established by means 
of dispersion theory.

\section{Unitarity and Resonances}

\subsection{Elastic unitarity}

Up to now we have not considered possible resonant states.
Resonances are one of the most characteristic features
of strong interactions. In our case, we expect them to appear at 
the 1 TeV scale. For instance, the MSM
becomes strong when $M_H\simeq1\mbox{TeV}$. In such case we expect a 
very broad scalar resonance around 1 TeV. In QCD-like models one 
expects a vector resonance around 2 TeV. 

From now on it will be very convenient
to use amplitudes of definite angular momentum $J$.
As far as we also have a conserved $SU(2)_{L+R}$ symmetry in
the EWSBS, we can also define a weak isospin $I$. In analogy
to $\pi\pi$ scattering, we will then have three possible isospin
channels $I=0,1,2$. At low energies we are only interested in
the lowest $J$, and thus we will concentrate
on the $t_{IJ}=t_{00},t_{11}$ and $t_{20}$ partial waves.
Indeed we will present our results in terms of their complex phases,
which are know as phase shifts.

Chiral lagrangians by themselves are not able to reproduce resonances.
Their amplitudes are obtained as polynomials in the momenta and masses,
and therefore they do not even satisfy the elastic unitarity condition:
\begin{equation}
\mbox{Im} t_{IJ}(s) = \sigma(s) \vert t_{IJ}(s) \vert^2
\label{uni}
\end{equation}
where $\sigma(s)$ is the two body phase-space.
Nevertheless, they satisfy it {\em perturbatively}
\begin{equation}
\mbox{Im} t^{(1)}_{IJ}(s) = \sigma(s) \vert t^{(0)}_{IJ}(s) \vert^2
\label{pertuni}
\end{equation}

Resonances are closely related to the
saturation of unitarity. That is why we have to unitarize
the chiral amplitudes.
There are many procedures in the literature to impose
Eq.\ref{uni} which very often lead to different results.
Obviously, that is one of the main criticisms to
unitarization. 

There is, however, a method that has been tested in ChPT
and is able to reproduce the $\rho$ and
$K^*$ resonances. It is based on dispersion theory
and apart from satisfying Eq.\ref{uni}, it also
provides the correct unitarity cut on the complex $s$ plane, as
well as poles in the second Riemann sheet.

\subsection{The inverse amplitude method}

If we consider an amplitude in the complex $s$ plane,
the existence of a threshold is reflected as a cut in the
real positive axis. The amplitude has two Riemann sheets
that are connected through the cut. 
By crossing symmetry, there is also another cut
on the left real axis. 

A dispersion relation is nothing but the Cauchy theorem applied
in one of the sheets. Thus, the values of that function in any point
will be given by the integrals of $\mbox{Im}t(s)$ over the cuts. 
Of course, these values are not known exactly, and 
with our chiral expansion we only get a crude approximation
replacing $\mbox{Im}t(s)\simeq\mbox{Im}t^{(1)}(s)$

The relevant point is to realize
that the inverse amplitude can be calculated {\em exactly} on the
{\em elastic cut}. 
Indeed, using Eqs.\ref{uni} and \ref{pertuni}
we find
\begin{equation}
\mbox{Im}\frac{1}{t_{IJ}}
=-\frac{\mbox{Im} t_{IJ}}{\mid t_{IJ} \mid ^2}
=-\sigma
=-\frac{\mbox{Im} t_{IJ}^{(1)}}{\mid t^{(0)}_{IJ} \mid ^2}
\end{equation}

Apart from poles, the cut structure of the amplitude $t(s)$ and
that of the function $\mid \mbox{t}^{(0)}_{IJ} \mid ^2/t_{IJ}(s)$
are the same. Their right cut contributions only differ on a sign,
and therefore, solving for $t(s)$ one obtains \cite{Truong,IAM}:
\begin{equation}
t_{IJ}\simeq\frac{t^{(0)}_{IJ}}{1-t_{IJ}^{(1)}/t^{(0)}_{IJ}}
\label{Pade}
\end{equation}
Notice that if we expand again at small momenta,
we recover the chiral expansion in Eq.\ref{trunc}.
Therefore, the Inverse Amplitude Method
(IAM) displays the correct low energy behavior.
We can perform again the very same analysis of
the preceeding section.
The difference from ChPT appears at higher energies, but
now we have several advantages:
\begin{Itemize}
\item It satisfies the elastic unitarity constraint.
\item The elastic right cut has been calculated exactly.
\item It can reproduce poles.
\end{Itemize}
Remember that the amplitude is extended continuously 
to the second Riemann sheet through the cut.
Hence, from the second point above,
we expect to obtain a very good approximation 
near the cut in the second Riemann sheet. 
But resonances are characterized as poles close
to the real axis and in the second sheet.
That is why this method is able to reproduce resonances.

Of course, the method has several limitations too \cite{IAM}.
First, the left cut is still an approximation. 
Next, we have neglected possible poles in G, which are indeed
present \cite{Penn}. Fortunately these effects are not dominant
at high energies, where the right cut and
resonance contributions dominate. They will however
introduce some uncertainty in the 
position and width of the resonances. 
There are also other rather technical issues that we will not
address here \cite{IAM}.

Let us now review how the IAM works. 
We want to know how well it reproduces the 
high energy behavior using {\em only low energy data},
since that could be the situation at LHC.
In particular, we are interested on whether
we can establish the existence of resonances
even though they are not directly seen.

\subsection{The IAM in Chiral Perturbation Theory}

When it is applied to pion physics \cite{Pade,IAM},
the IAM is able to reproduce a 
$\rho$ resonance just using low energy data.
In Figure 1.a, the results of plain ChPT
are plotted as a dotted line.
It has been calculated with
the parameters proposed in \cite{Rigg},
which have been obtained {\em only} from
{\em low energy data} ($\lappeq$400 GeV).
The other two lines are the IAM prediction.
The dashed one has been obtained with the same parameters
and the continuous one with
an slightly different set \cite{BiGa}.
\begin{figure}
\vskip -1.cm
\leavevmode

\begin{center}
\hspace{-2mm}
\mbox{\epsfysize=5.1cm\epsffile{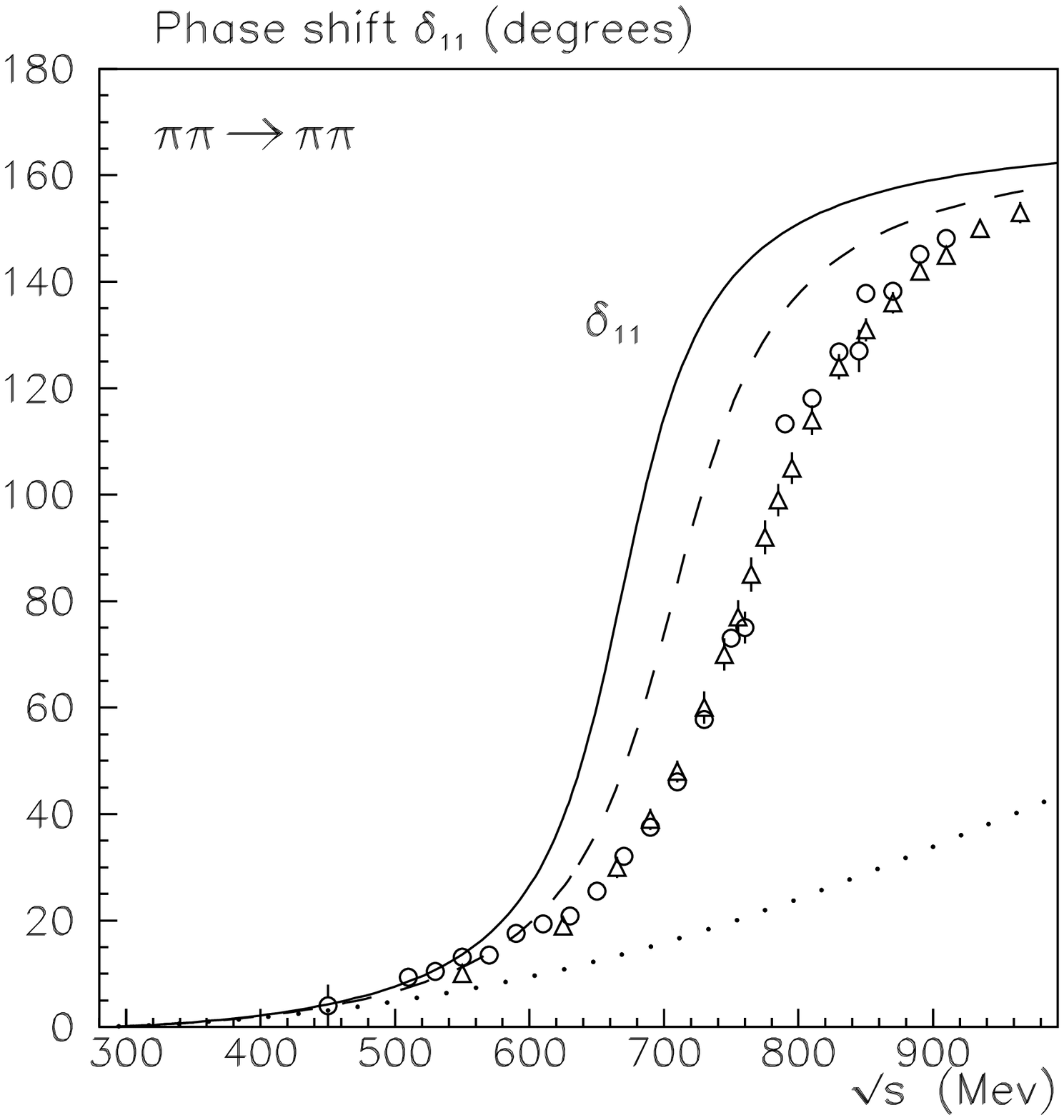}
\hspace{-1.cm}
\epsfysize=5.1cm\epsffile{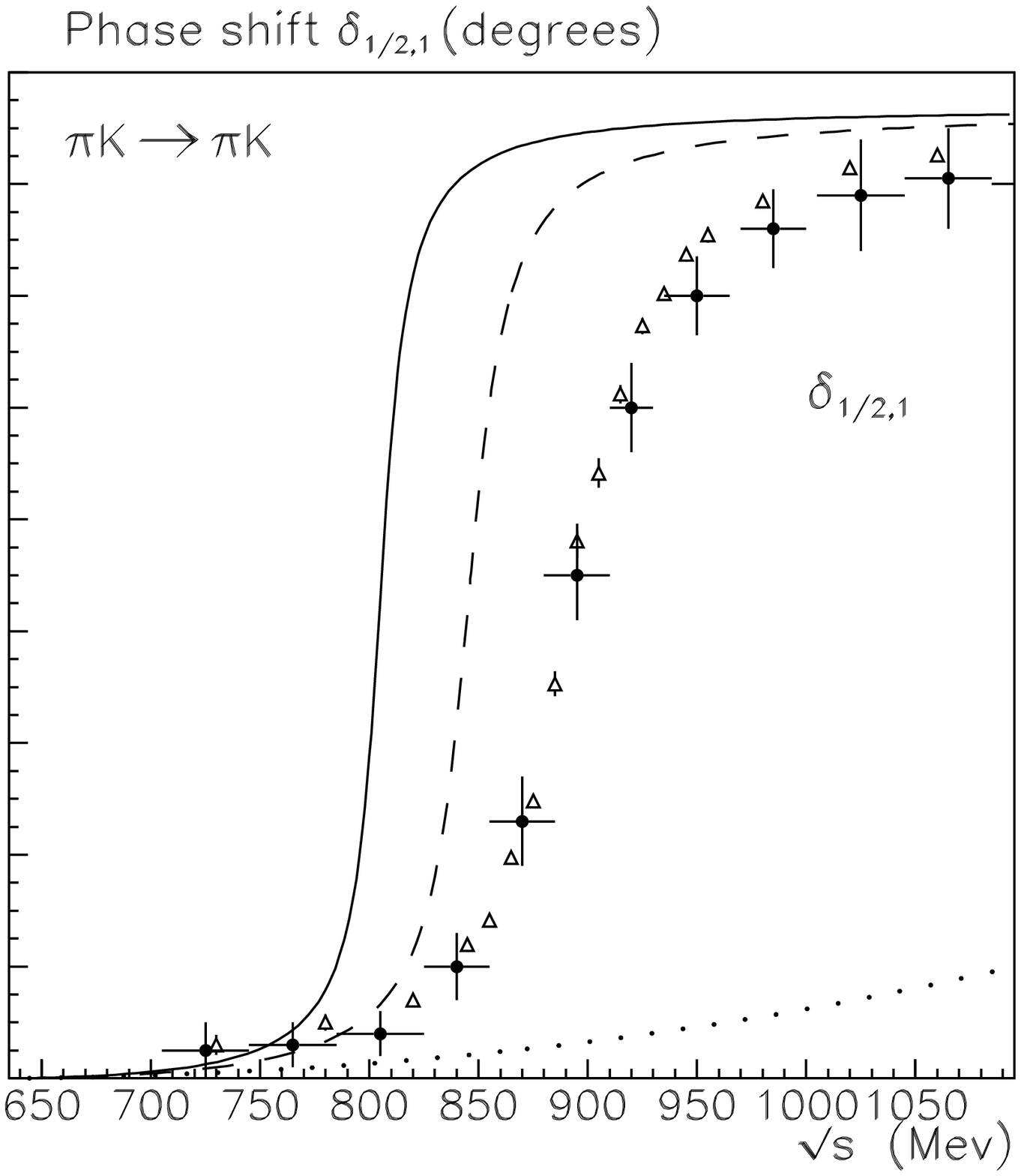}}
\end{center}

\vskip -.5cm

{\bf Figure 1.-}a) $\delta_{11}$ phase shift in $\pi\pi$
scattering. The data comes from: \cite{Proto} ($\triangle$),
\cite{Esta} ($\circ$);
b)$\delta_{1/2,1}$ in $\pi K$ scattering.
\cite{MeAn71} ($\bullet$),\cite{EsCa78} ($\triangle$).
The dotted curves are plain ChPT. The
others are the IAM with two sets of chiral parameters.
\end{figure}
As far as the only input in the calculations
is low energy data, the existence of the $\rho$
can be seen as a {\em prediction} of the IAM.
The qualitative behavior of the phase shift
is obviously correct. Notice that
the value of its mass does not lie very far 
from the actual value. The theoretical error
is hard to estimate, but we have found, varying
the parameters inside their error bars, that
it is never bigger than $20\%$ \cite{IAM}. 

Of course, it is possible to get a better fit (see
\cite{IAM}) but then high energy data should also be 
used as an input. 

The case of $\pi\pi$ scattering is specially relevant
since it can be described 
with the very same $SU(2)$ scheme of symmetry breaking
of the EWSBS.
However, the IAM also works in other
models.  In Figure 1.b it is shown
how it is also possible to reproduce the $K^*(892)$ resonance in
$\pi K$ elastic scattering using $SU(3)$ ChPT \cite{Pade,IAM}.
The uncertainties are again of the same order.

It can also be checked \cite{IAM} that the amplitudes
present the appropriate analytical structure including
the corresponding poles in the second Riemann sheet.

We have therefore shown that the IAM is
not just a simple numerical trick to unitarize amplitudes.
It contains all the analytic structure needed to
extract the correct high energy behavior
from low energy data. 

\subsection{Resonances in the strong EWSBS.}

Throughout this section we will be using the
Equivalence Theorem \cite{ET}. It states that the
$V_LV_L$ amplitudes are those of GB up to $\Od(M_V/\sqrt{s})$.
At high energies those terms can be neglected and
the $V_LV_L$ amplitudes look exactly as those
of $\pi\pi$ scattering in the massless limit.

At first sight it is not evident that such a high energy
limit can be used with a low energy approach like
chiral lagrangians. However, it has been shown \cite{NPB,chinos}
that there is a common applicability window, and that
the theorem remains the same when working at lowest
order in the electroweak couplings, which is our case.

Let us then apply the IAM to
the reference models of Table I.
In Figure 2 we can see (solid lines)
how the IAM yields an scalar
resonance in the Higgs model, and a technirho
in the QCD model \cite{DoHeTe}. 
There are no other resonances present.
\begin{figure}

\vskip -1.cm

\leavevmode
\begin{center}
\mbox{\epsfysize=5.cm\epsffile{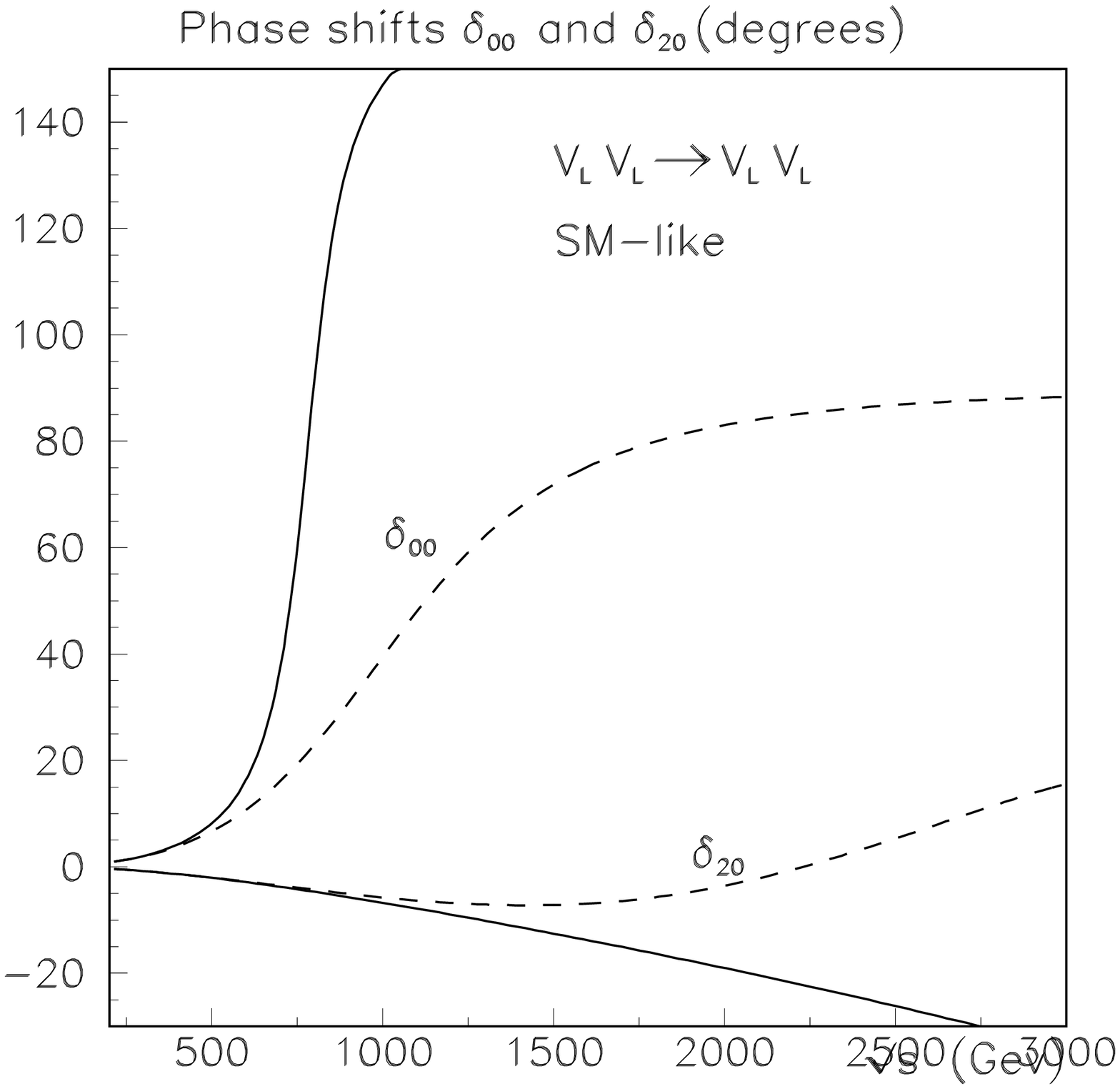}
\hspace{-6mm}
\epsfysize=5.cm\epsffile{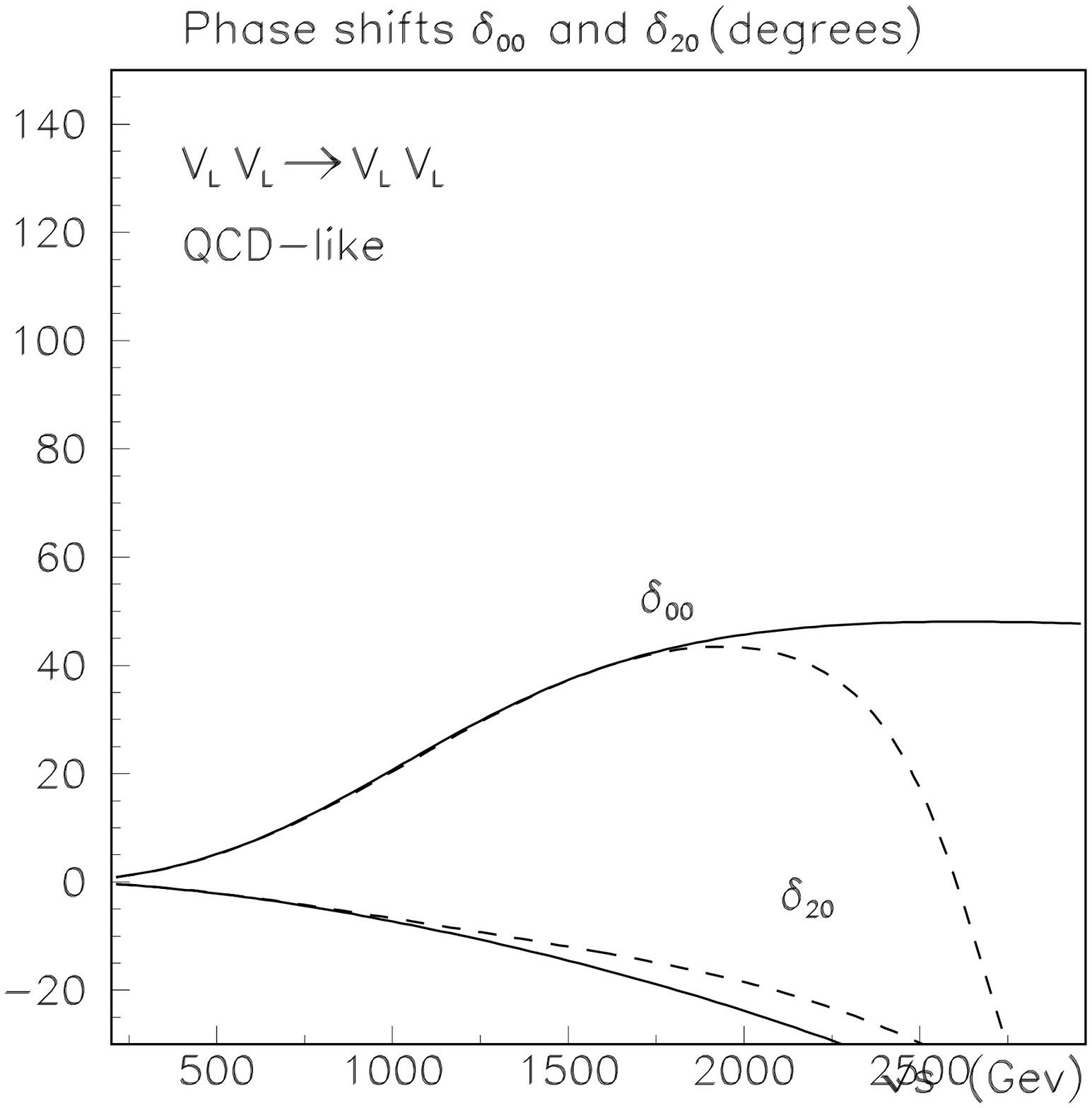}}
\mbox{\epsfysize=5.cm\epsffile{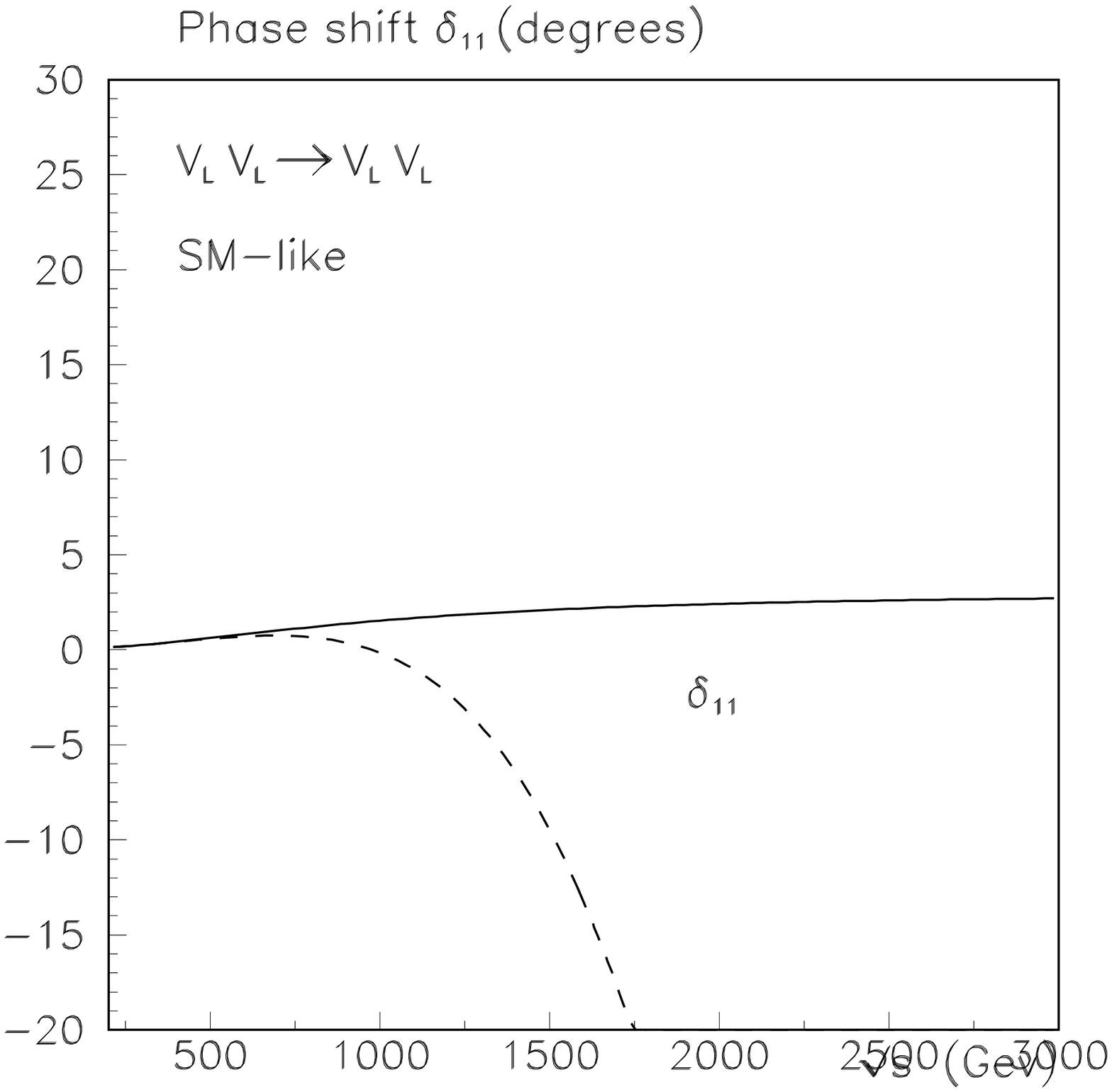}
\hspace{-6mm}
\epsfysize=5.cm\epsffile{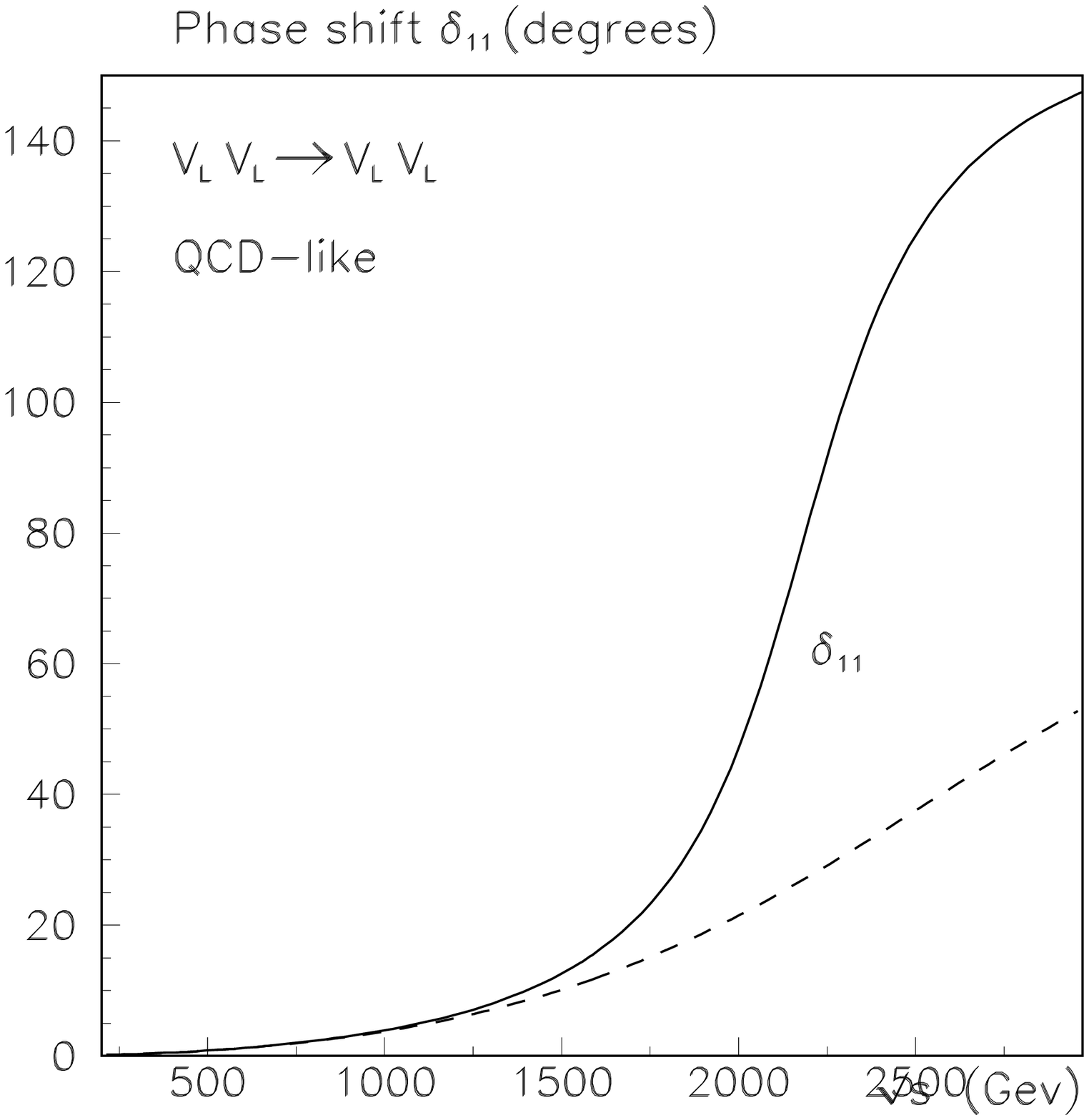}}
\end{center}

\vskip -.5cm

{\bf Figure 2.-} 
$V_LV_L\rightarrow V_LV_L$ phase shifts in the heavy Higgs SM
(left) and a QCD-like model(right). Notice their respective
 scalar and the vector resonances.
The dashed lines are the chiral amplitudes
and the solid lines are the IAM results. 

\end{figure}
We have found again that the IAM yields 
the correct result. Let us then scan the parameter space
to get a qualitative description of the
general resonance spectrum of an strong EWSBS.

We will only concentrate on the $(I,J)=(0,0)$
and $(1,1)$ channels. The $I=2$ channel is more subtle
and will be given elsewhere.

\begin{figure}

\vskip -1.cm

\leavevmode
\begin{center}
\mbox{\epsfysize=5.1cm\epsffile{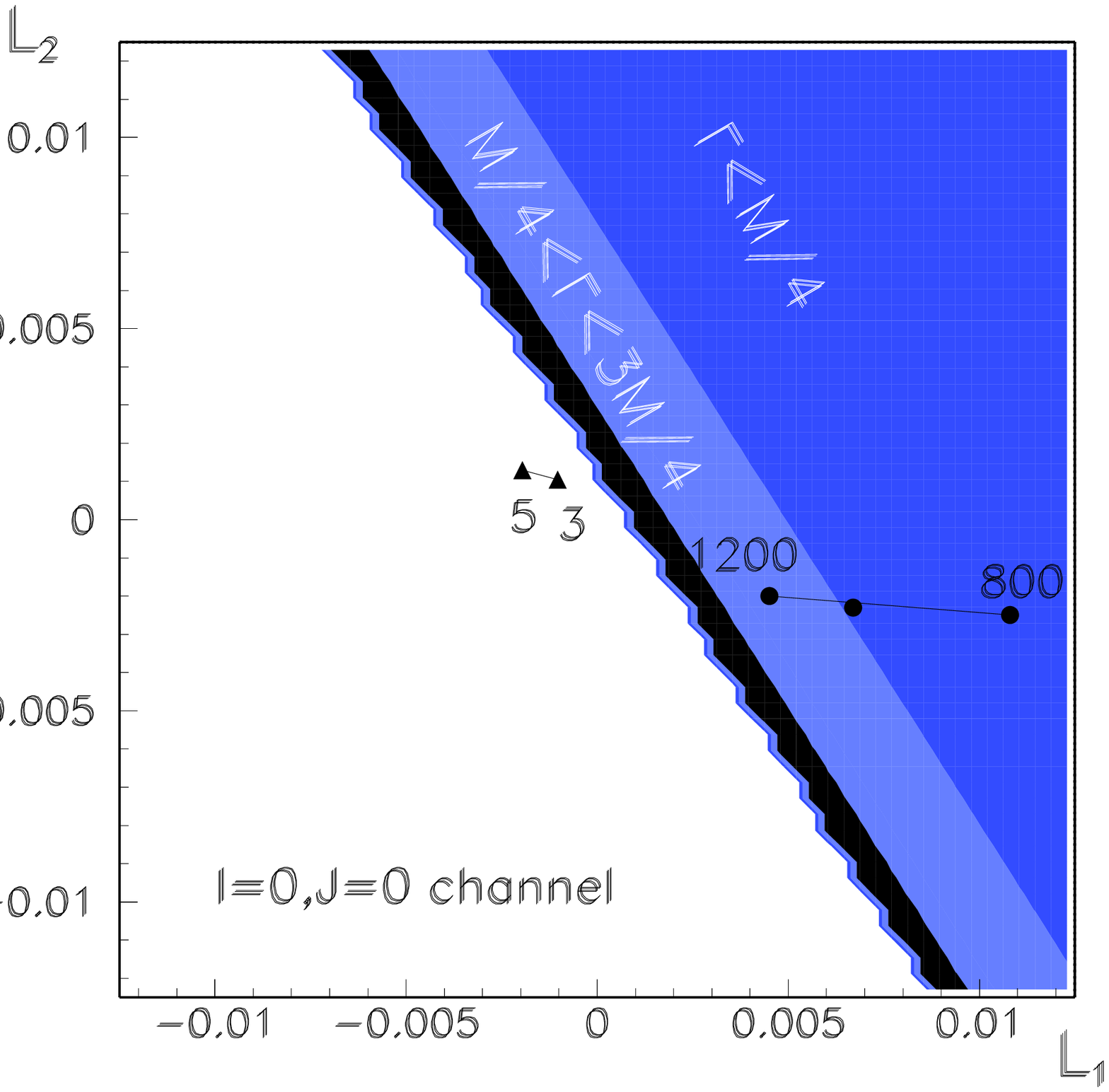}
\hspace{-1cm}
\epsfysize=5.1cm\epsffile{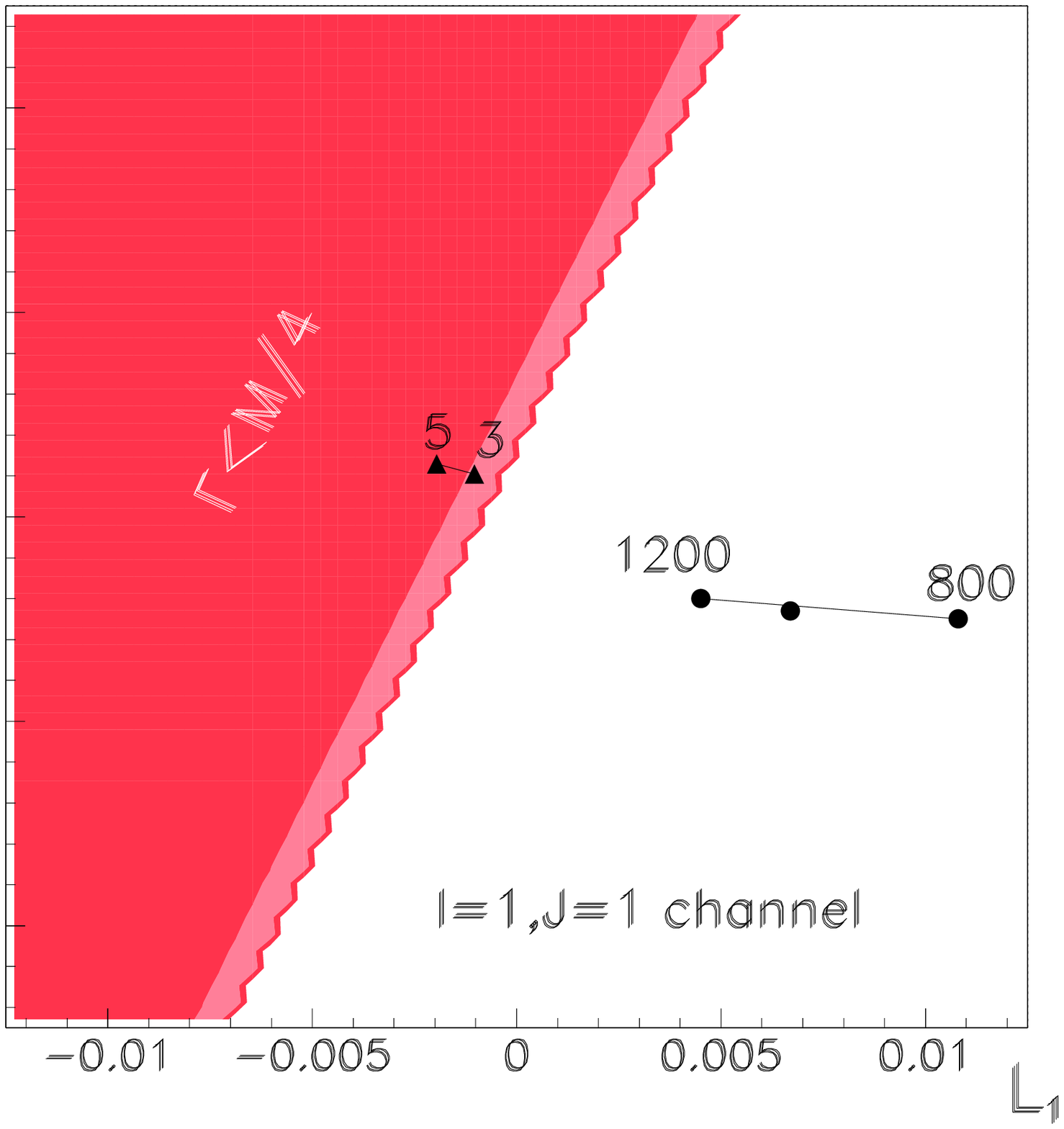}}
\end{center}

\vskip -.5cm

{\bf Figure 3.-} Resonant states in the
$L_1,L_2$ plane, both for the $(I,J)=(0,0)$
and $(1,1)$ channels. The dark color areas
correspond to narrow resonances. Lighter
areas are broad resonances and black areas
stand for saturation. White is no resonance
or saturation below $3 \mbox{TeV}$

\end{figure}

In Figure III we have plotted in the 
$L_1, L_2$ plane the expected unitarity
behavior up to 3 TeV of the $VV$ amplitudes.
There are several possibilities: No resonance (white),
a saturation of unitarity (black), a broad resonance
(light) or a narrow resonance (dark). By narrow
or broad, we mean that the width is smaller or
bigger than 25\% of the mass, respectively.
We understand by saturation that the 
unitarity bound is reached, but a resonance there would
have a width of 75\% its mass or more.
We have also shown the position
 of the SM with $M_H=800$ to $1200\mbox{GeV}$
(black dots),
as well as QCD-like models with 3 or 5 technicolors
(black triangles).

From the graphs it seems that there are many
different phenomenological scenarios. Maybe there is
just one resonance, two resonances or no resonances at
all. It could happen that one channel saturates unitarity
while the other has a resonance, etc...

In conclusion,
the effective lagrangian approach supplemented with
the IAM, emerges as a very powerful
and simple tool to explore
a great variety of strongly interacting scenarios.

%

\end{document}